\documentclass[fonts]{icst}

\usepackage{moreverb}

\usepackage[breaklinks,colorlinks,bookmarksopen,bookmarksnumbered,linkcolor=ICSTblue,citecolor=blue,urlcolor=ICSTblue]{hyperref}
\usepackage{breakurl}

\usepackage{amssymb}
\usepackage{graphicx}
\usepackage[ruled,vlined,linesnumbered]{algorithm2e}
\usepackage{float}
\usepackage{color}
\usepackage{soul}
\usepackage{amsmath}
\usepackage{caption}
\usepackage{makecell}

\usepackage{url}

\newcommand\BibTeX{{\rmfamily B\kern-.05em \textsc{i\kern-.025em b}\kern-.08em
		T\kern-.1667em\lower.7ex\hbox{E}\kern-.125emX}}

\def\volumeyear{2014}

\journalname{XXXXXX}
\articletype{Research Article/Editorial}
\def\volumeyear{XXXX}
\def\volumenumber{XX}
\def\issuemonth{XX-XX}
\def\issuenumber{X}
\def\articleid{XX}
\def\copyrightholder{Iqbal H. Sarker \textit{ et al.}, licensed to ICST}
\copyrightnote{This is an open access article distributed under the terms of the Creative Commons Attribution license (\url{http://creativecommons.org/licenses/by/3.0/}), which permits unlimited use, distribution and reproduction in any medium so long as the original work is properly cited.}
\def\articledoi{10.4108/XX.X.X.XX}
\received{XXXX}
\accepted{XXXX}
\published{XXXX}

\begin{document}

\runningheads{Iqbal H. Sarker}{SilentPhone: Inferring User Unavailability based Opportune Moments to Minimize Call Interruptions}

\title{SilentPhone: Inferring User Unavailability based Opportune Moments to Minimize Call Interruptions}

\author{Iqbal H. Sarker \affil{*}}

\address{Department of Computer Science and Software Engineering, \\ Swinburne University of Technology, \\ Melbourne, VIC-3122, Australia.}

\abstract{The increasing popularity of cell phones has made them the most personal and ubiquitous communication devices nowadays. Typically, the ringing notifications of mobile phones are used to inform the users about the incoming calls. However, the notifications of inappropriate incoming calls sometimes cause interruptions not only for the users but also the surrounding people. In this paper, we present a \textit{data-driven} approach to \textit{infer the opportune moments} for such phone call interruptions based on \textit{user's unavailability}, i.e., when a user is unable to answer the incoming phone calls, by analyzing individual's past phone log data, and to discover the corresponding phone \textit{silent mode configuring} rules for the purpose of minimizing call interruptions in an automated intelligent system. Experiments on the real mobile phone datasets show that our approach is able to identify the opportune moments for call interruptions and generates corresponding silent mode configuring rules by capturing the dominant behavior of individual users' at various times-of-the-day and days-of-the-week.}


\tnotetext[1]{This article is a preprint version of the journal \textit{"EAI Endorsed Transactions on Mobile Communications and Applications"}. \\}

\fnotetext[1]{Corresponding author: Iqbal H. Sarker.  Email: \email{msarker@swin.edu.au}}

\maketitle

\section{Introduction}
\label{sec1}
The growing adoption and the popularity of mobile phones have dramatically changed the way we interact and communicate with other people in the current world. The mobile phone represents highly personal device of an individual's daily life. These mobile phones are considered to be `always on, always connected' devices. However, the mobile phone users are not always attentive and responsive to incoming communications because of their day-to-day situations in their daily activities. For this reason, sometimes people are often interrupted by incoming phone calls which not only create disturbance for the owners but also for the people nearby. Such kind of interruptions may create embarrassing situation not only in an official environment (e.g., meeting) but also affect in other activities like examining patients by a doctor or driving a vehicle etc. Sometimes these kinds of interruptions may reduce worker performance, increased errors and stress in a working environment \cite{pejovic2014interruptme}.

Mobile phones have the obvious benefit of all the moment communication \cite{zulkernain2010mobile}. Typically, we can expect a mobile phone to ring to inform the users about the incoming calls. However, ringing phone at an inopportune moment can be very disruptive to the current task or social situation. The interruptions greatly impact on knowledge worker productivity. According to the Basex BusinessEdge report \cite{spira2005cost}, interruptions consume 28\% of the knowledge worker's day, which is based on surveys and interviews conducted by Basex over the 18 months period encompassing high-level knowledge workers, senior executives at the end-user organizations, and executives at companies that produce Collaborative Business Knowledge tools. This translates into 28 billion lost man-hours per annum to companies in the United States alone. It results in a loss of \$700 billion, considering an average  salary of \$25/hour for a knowledge worker, according to Bureau of Labor Statistics \cite{laborStatistics}. In another study related to the execution time of primary tasks, Bailey et al. \cite{bailey2006need} have shown that when interrupted users require from 3\% to 27\% more time to complete the tasks, and commit twice the number of errors across the tasks. Thus managing phone call interruptions is one of the key research areas in the mobile phone domain.

A number of authors \cite{dekel2009minimizing} \cite{khalil2005improving} \cite{zulkernain2010mobile} have studied about the phone call interruption management systems. However, the main drawback of these systems is that the phone ringer mode configuring rules used by the applications are not automatically discovered; users need to define and maintain the rules manually by themselves. In contrast, calling activity records in device logs are a rich resource for analyzing the phone call activity patterns of individuals \cite{sarker2016phone}. In this paper, we aim to discover \textit{temporal phone silent mode configuring rules} based on user's unavailability utilizing individual's phone log data, i.e., the opportune moments to minimize call interruptions.

Mobile phones automatically record the users activities related to phone calls in it's log. Their ability to log call activities offers the potential to understand the call response behavior of individuals. In particular the storing of calling information (e.g., the temporal information, calling activities, and others call related meta-data) in call logs provides raw contextual information about when the user is unable to answer the incoming calls. The temporal patterns of these responses may provide the basis for inferring the opportune moments for interruptions based on \textit{user unavailability}. Therefore, our main objective in this paper is - inferring the \textit{opportune moments} for interruptions based on \textit{user's unavailability} and to generate the corresponding \textit{silent mode configuring rules} utilizing their own phone log data. Based on these rules, an intelligent interruption management system can be developed for the end mobile phone users to automatically configure silent mode for the purpose of minimizing call interruptions in their daily activities.

In particular, the contributions of this paper are summarized as below: 

\begin{itemize}
	\item We highlight the importance of minimizing phone call interruptions in our daily activities.
	
	\item We infer the opportune moments to minimize call interruptions based on user unavailability by capturing the dominant behavior of individuals at various times-of-the-day and days-of-the-week
	
	\item Our experiments on real-life mobile phone datasets show that our approach is effective to identify the opportune moments for call interruptions and to generate corresponding silent mode configuring rules.
\end{itemize}

The rest of the paper is organized as follows. Section \ref{Related Work} reviews the related work. Section \ref{User-Unavailability with Phone Calls} briefly discusses user unavailability related to phone calls. We present our approach in Section \ref{Our Approach}. Section \ref{Experiments} reports some experimental results on mobile phone datasets and finally Section \ref{Conclusion and Future Work} concludes this paper and highlights the future work.

\section{Related Work}
\label{Related Work}
A significant amount of research has been done on user survey about phone call interruptions and their management systems. For instance, Khalil et al. \cite{khalil2006context}, have shown the usefulness of interruption management system by conducting a user survey. In their survey, they investigate context disclosure and sharing patterns for context-aware telephony with the aim of decreasing interruptions and enhancing agreement between callers and receivers. They found the low availability rate (only 53\% of the time) of the participants to receive cell phone calls. In another work, Toninelli et al. \cite{toninelli2009s} have reported a survey of activity-based response to incoming calls of different users and show that maximum users do not want to interrupt while in meeting or working in a team or outdoor activities like driving or sleeping and ignore incoming calls in these situations.

For the purpose of minimizing such interruptions, Pejovic et al. \cite{pejovic2014interruptme}, design and implement an interruption management library based on a number of sensors of Android phones. In \cite{smith2014ringlearn}, the author presents a new approach to smartphone interruptions that maintains the quality of mitigation under concept drift with long-term usability. The approach uses online machine learning and gathers labels for interrupt causing events (e.g., incoming calls) using implicit experience sampling without requiring extra cognitive load on the user's behalf. Another research is quietly different based on UI (User Interface). Authors present a multiplex UI for handling incoming calls on smartphones \cite{bohmer2014interrupted}. This design solution tackles the problem that calls can interrupt concurrent application uses. They extended the options for handling incoming phone calls and presented considerations for possibilities to postpone calls and multiplex the call notification with the concurrent application.

Besides these, a number of authors have studied about the rule-based systems that can manage these interruptions. However, such research does not take into account configuring rules produced by automatically analyzing phone log data. For instance, Khalil et al. \cite{khalil2005improving} use calendar information to automatically configure cell phones accordingly to manage interruptions. Seo et al. \cite{seo2011pyp} propose a context-aware configuration manager for smartphones to block a phone call without bothering the user. Dekel et al. \cite{dekel2009minimizing} design an application to minimize mobile phone disruptions. An intelligent context-aware interruption management system has designed by Zulkernain \cite{zulkernain2010mobile}. The main drawback of these approaches is that these are not data-driven, i.e., the configuring rules used by the applications are static and not individualized. A set of automated discovered rules according to individual's phone call activity patterns can make the interruption management system more effective and personalized, in which we are interested in.

Unlike the above works, we utilize phone log data of individuals and infer the opportune moments based on their unavailability with phone calls by analyzing their past phone call activities.

\section{User-Unavailability with Phone Calls}
\label{User-Unavailability with Phone Calls}
In the real world, the common phone call activities of an individual mobile phone user are - (i) answering the incoming phone call by the user for a particular time period or duration, i.e., `Accept', (ii) instantly decline the incoming phone call by the user, i.e., `Reject', (iii) the phone rings for an incoming call but the user misses the call, i.e., Missed, and (iv) making a phone call to a particular person, i.e., `Outgoing'. Except the outgoing call, all the calling activities are related to incoming communications from another person. However, the users are not always attentive and responsive to incoming phone calls in their daily activities in the real world. Because of their various day-to-day situations, sometimes they are unavailable to answer the incoming phone calls and as a result causes interruptions. According to the above definition, making an outgoing call is not related to the user unavailability or call interruptions, as the user him/her self initiates such calls. As such, we ignore the outgoing call in our unavailability analysis in this work.

Identifying user unavailability for answering the incoming phone calls, is the key of our approach. A number of researchers assume user unavailability based on calendar schedules or events. According to Khalil et al. \cite{khalil2005context}, calendar entries can be a good cue as to whether a person is available or unavailable for a phone call. For instance, if the calendar has a meeting appointment between 13:00 and 14:00, they assume with a high degree of probability that the user is unavailable and she is in a place with at least one other person. Salovaara et al. \cite{salovaara2011phone} have conducted a study and show that 31\% of the incoming phone calls were unavailability related, i.e., the users are unavailable to answer the phone calls for various events such as meetings, lectures, appointments, driving, sleeping. However, such unavailability solution provides low accuracy in some cases because of lacking the behavioral evidence of individuals \cite{sarker2016evidence}. Khalil et al. \cite{khalil2005improving} surveyed 72 phone users and found that the above unavailability solution for mobile communication gives low accuracy (62\%) for loosely structured home activities such as `lunch', `watching TV', `homework', while gives high level of accuracy (93\%) for the structured events such as `lectures', `meetings', `appointments'. However, such special words are not sufficient for the real life use cases to cover, need a richer set of meeting categories keywords to capture the actual behavior of the users \cite{dekel2009minimizing}.

The above event-based user unavailability having a static temporal period is based on \textit{assumptions} and \textit{not individualized}. According to \cite{pielot2014large}, the common reasons that make the users unavailable to answer an incoming call are being in a meeting or work, or prefer to focus on other activities, such as sleeping or playing games. In that situations, the incoming calls are either \textit{rejected} by the users directly or \textit{missed}, that represents the ``User-Unavailability'' for phone calls. Thus, in this paper, we consider the user is unavailable if s/he does not answer the incoming calls, in other words, if the user rejects or misses the incoming calls. Therefore, by taking into account such user unavailability, in this work, we aim to infer the opportune moments according to their own preferences for minimizing phone call interruptions, by analyzing the phone call activities recorded in their mobile phone log.

\section{Our Approach}
\label{Our Approach}
In this section, we present our approach to infer the opportune moments based on user unavailability discussed above by analyzing individuals phone log data, in order to minimize the interruptions. Our approach accepts as input a real phone log dataset of an individual user consisting of phone call activities, temporal information, and other call related meta-data. In order to achieve our goal, we first pre-process the log dataset for the purpose of getting useful patterns according to our interests in this work.

\subsection{Log Data Pre-processing} 
As discussed above, the call unavailability is related to reject or missed call. However, the mobile phone automatically records reject call as an incoming call with call duration zero (0 second) in the device log \cite{eagle2006reality} \cite{sarker2016behavior}. An incoming call that has call duration more than zero, is differentiated as the accept call, i.e., the user answered the incoming calls for a particular time period \cite{sarker2016behavior}. Table \ref{sample-log-data} shows an example of these call activities captured by the mobile device, with corresponding aspects including user unavailability with phone calls according to log data.

\begin{table*}[htbp!]
	\begin{center}
		\caption{Sample call details with unavailability}
		
		\label{sample-log-data}
		\begin{tabular}{l|c|c|c} 
			\hline
			\makecell{\textbf{Aspects}} & \makecell{Call Accepted \\ by User} & \makecell{Call Rejected \\ by User} & \makecell{Call Missed \\ by User}\\
			\hline
			\makecell{Call Type} & \makecell{INCOMING} & \makecell{INCOMING} & \makecell{MISSED}\\
			\hline
			\makecell{Call Duration} & \makecell{$>0$ (zero) second} & \makecell{$=0$ (zero) second} & \makecell{$=0$ (zero) second}\\
			\hline
			An example & \makecell{Date: 2015-04-12 \\ Time: 12:10:20 \\ Call Type: INCOMING \\ Call Duration: 125 sec} & \makecell{Date: 2015-04-13 \\ Time: 12:30:55 \\ Call Type: INCOMING \\ Call Duration: 0 sec} & \makecell{Date: 2015-04-14 \\ Time: 12:50:20 \\ Call Type: MISSED \\ Call Duration: 0 sec}\\
			\hline
			\makecell{Unavailability} & \makecell{No} & \makecell{Yes} & \makecell{Yes}\\
			\hline
		\end{tabular}
	\end{center}
\end{table*}

\subsection{Extracting Temporal Patterns of Unavailability} 
As can be seen in Table \ref{sample-log-data}, phone log contains the exact time (e.g., YYYY-MM-DD hh:mm:ss) for each call activity of individual users. However, human understanding of time is not precise, unlike digital systems. Our daily behavioral activities occur in time intervals rather than an exact time \cite{sarker2017individualized}. For instance, an employee does not attend in a meeting every day at exactly the same time, or drives to work exactly the same time every day. There is always a time interval for routine behaviors, even if only a small interval, e.g., five minutes. Therefore, there is a need for temporal granularity to capture the patterns of individual's unavailability to answer the incoming phone calls in their daily activities.

To do this we follow the bottom-up approach utilizing individual's phone log data. First, each day (24-hours-a-day) of the week is divided into relatively small time slices (e.g., Monday[10:00-10:10]) using a base time period (say, 10 minutes). \textit{`A base period is taken into account as the smallest unit of time period to capture the \textit{user unavailability patterns} according to her phone call activities in various day-to-day situations, recorded in her phone log'}. 

In order to determine the unavailability period, we then identify the dominant \cite{sarker2017individualized} call activity of each slice according to a confidence threshold (say, $t = 80\%$) preferred by an individual. In a time slice, if the percentage of reject ($R$) or missed ($M$) calls separately or combinedly ($R$ and $M$) of a user is greater than $t$, the corresponding time slice will be considered an unavailability period for that user. After identifying such periods during the whole 24-hours-a-day time period, we aggregate the adjacent unavailability periods in order to get more meaningful or larger period according to the similar characteristics. Such periods can be used to generate silent mode configuring rules for the users. If consecutive slices with unavailability dominant are not found, then the individual slice with unavailability dominant can be used as a single unavailability period and is allowed to generate silent mode configuring rules as well. Thus aggregating or merging operation is only applied for consecutive time slices with the dominant of unavailability. This process has been set out in Algorithm \ref{alg:unavailability-generation}.

\begin{algorithm}
	\caption{Unavailability Period Generation}
	\label{alg:unavailability-generation}
	\SetKwInOut{Data}{Data}
	\SetAlgoLined
	\Data{Dataset: $DS = {X_1,X_2,...,X_n}$ // each instance $X_i$ includes phone call activity and corresponding temporal information.}
	
	\KwResult{A list of unavailability periods $UN_{list}$}
	
	\BlankLine
	
	$TS_{list} \leftarrow generateTimeSlices()$; // generate initial time slices
	using a base period during the whole 24-hours-a-day. \\
	
	\ForEach{slice $TS \in TS_{list}$}
	{  
		$UN_{dom} \leftarrow identifyUnavailDominant(TS)$; //identify unavailability dominant according to the activity occurances in a slice.\\
		
		//check whether unavailability dominant found or not in the slice \\
		\If{$UN_{dom}$ is true}
		{
			$UN_{new} \leftarrow createUnavail (TS);$ // create a new unavailability period using the current time slice. \\
			store $UN_{new}$ in the list $UN_{list};$ \\
			//check for consequtive slices with unavailability dominants \\
			\If{consequtive $UN_{dom}$ exist}
			{
				$UN_{updated} \leftarrow mergeSlices (current, next);$ // merge the consequtive time slices and update the time boundaries accordingly. \\
				// update the unavailability list. \\
				updateList ($UN_{new}$, $UN_{updated}$);
			}
		}		
	} 
	
	return $UN_{list}$ \\ 	
\end{algorithm}

As we do not assume any prior knowledge about the optimal base period and corresponding time slices, we therefore generate a number of sets of configuring rules using the merged unavailability periods generated using Algorithm \ref{alg:unavailability-generation} and associated days, by iteratively varying the base time period. In addition to the temporal period, we also take into account the day-wise phone call activity and corresponding user unavailability period while determining the optimal period. The reason is that individuals daily activities may vary from day-to-day and the unavailability periods as well as the unavailability to answer the phone call is related to individuals various day-to-day situations in their daily activities. For instance, one's Wednesday's activities and corresponding unavailability periods may not be the similar with her Thursday's activities and unavailability periods. 

To determine the optimal period, we measure the applicability of the generated rules sets for each base time period. If, $Rsup$ represents the support count of a rule, $Rcov$ represents the temporal coverage of the rule, $S_{max}$ represents the maximum possible support in a dataset, $C_{max}$ represents the maximum possible temporal coverage in a week, and `N' is the number of rules that satisfies the user's confidence threshold, then according to \cite{sarker2016behavior}, the applicability is defined as below:

\begin{equation}
	\label{applicability}
	Applicability =\sum_{i=1}^N \left( \frac{Rsup_i}{S_{max}} * \frac{Rcov_i}{C_{max}} \right) 
\end{equation}

The base time period (say, 20 minutes) and corresponding time slices that yields maximum applicability establishes the optimal period for capturing best patterns of individuals unavailability periods according to their reject and missed call activities in their own log data. This optimal base time period may vary from day-to-day and from user-to-user. The reason is that different users have different phone call activity patterns and time periods in different days-of-the-week (Monday, Tuesday,...,Sunday) and therefore different unavailability periods to answer the phone calls according to their own preferences.

\subsection{Generating SILENT Mode Configuring Rules} 
To minimize the phone call interruptions, the ``SILENT'' mode should be configured automatically as the phone ringing mode. The reason is that silent is a common ringer mode of mobile phones for controlling signals of notifications during interruptions, which is available in every mobile phones. Therefore, to discover the silent rules according to the above unavailability time periods, we employ the well-known association rule mining algorithm Apriori \cite{agrawal1994fast}. The reason is that association-rule learning is a well-defined, and allows to configure the preferences of individuals in terms of support and confidence, in which we are interested in. If  $X \rightarrow Y$ is a SILENT mode configuring rule, where X represents the unavailability time period and Y is the phone SILENT mode, then according to \cite{agrawal1994fast} -

\begin{itemize}
	\item \textit{Support}: the ratio between the number of times X and Y co-occur and the number of data-instances present in the given data. It can be represented as the joint probability of X and Y : $P(X,Y)$.
	
	\item \textit{Confidence}: the ratio between the number of times Y co-occurs with X and the number of times X occurs in the given data. It can be represented as the conditional probability of X and Y : $P (Y | X)$.
\end{itemize}

Such preferences and corresponding SILENT mode configuring rules may differ from user-to-user as to how interventionist they want the call-handling agent to be. For instance, one individual may want the automated response handling agent to configure the mobile phone silent, where in the past he/she has rejected or missed calls more than, say, 80\% of the time - that is, confidence threshold of 80\%. Another individual, on the other hand, may only want the agent to intervene if he/she has rejected or missed calls in, say, 100\% of past instances.

\section{Experiments}
\label{Experiments}
In this section, we have conducted experiments on the real mobile phone datasets of individual mobile phone users. We have implemented our approach in Java programming language and executed them on a Windows PC. In the following subsections, we briefly describe the datasets, and present the experimental results.

\subsection{Mobile Phone Dataset} 
We have conducted the experiments on the real mobile phone dataset (`Swin') that was collected by us from 22 individual mobile users, such as undergraduate students, post graduate students, university lecturers and industry professionals, from August 2014 to September 2015, at Swinburne University of Technology, Melbourne, Australia \cite{sarker2016behavior}. In the following, we report the average experimental results utilizing these phone log datasets, and the detailed of experimental results of two individual users, represented as U1 and U2.

\subsection{Produced SILENT Mode Configuring Rules} 
Using our approach, we get a set of silent mode configuring rules for each individual user according to her own activity patterns utilizing her own phone log dataset. The followings are the examples of discovered rules for a particular user U1 having preference of 80\% confidence.

\begin{eqnarray*}
	R1: DayTime \rightarrow Friday [16:15-17:30] \\ 
	\Rightarrow RingerMode  \rightarrow Silent \; (Conf=100\%) 
\end{eqnarray*}
\begin{eqnarray*}
	R2: DayTime \rightarrow Monday [10:00-11:40] \\ 
	\Rightarrow RingerMode  \rightarrow Silent \; (Conf=87\%)
\end{eqnarray*}
\begin{eqnarray*}
	R3: DayTime \rightarrow Monday [14:20-16:00] \\ 
	\Rightarrow RingerMode  \rightarrow Silent \; (Conf=92\%)
\end{eqnarray*}

Rule R1 states the user is always (100\%) interrupted between 16:15 and 17:30 on Fridays, and according to this rule the user U1 needs to keep her phone silent in this time period in order to minimize call interruptions on this particular day. Similarly, the rule R2 and R3 state that the user is also interrupted between 10:00 and 11:40, and between 14:20 and 16:00 on Mondays, most of the time 87\% and 92\% respectively, and needs to keep her phone silent in these time periods as well to minimize interruptions on this particular day. These silent notification mode rules are discovered based on the patterns of reject and missed call activities recorded in her own phone log dataset. Similarly, for another user U2, another set of such rules are discovered based on her own unavailability periods recorded in her own dataset.

\begin{figure}[htbp]
	\centering
	\includegraphics[width=\linewidth, keepaspectratio]{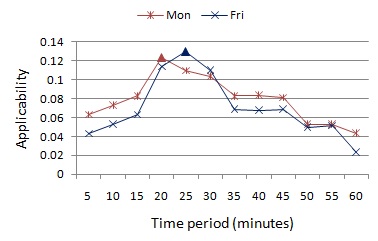}
	\caption{Optimal base time period selection}
	\label{fig:base-period}
\end{figure}

\begin{figure}[htbp]
	\centering
	\includegraphics[width=\linewidth, keepaspectratio]{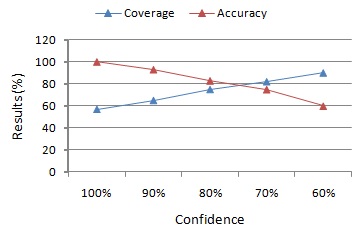}
	\caption{Avg. results (accuracy and coverage) utilizing a collection of datasets for various confidence preferences}
	\label{fig:Dataset-2}
\end{figure} 

These rules are generated based on the optimal time period discussed in the earlier section, which is shown in Figure \ref{fig:base-period} (up to 60 minutes) for different days Monday and Friday respectively, for a particular confidence preference 80\% of User U1. The base period that produces the highest (peak) applicability, is the \textit{optimal base period}. From Figure \ref{fig:base-period}, we found that 20 minutes (Monday) and 25 minutes (Friday) are the optimal base periods for generating time slices and corresponding unavailability periods on these days for this user. For another user U2, this optimal base period may differ according to her activities and preference, and corresponding unavailability periods are generated accordingly.

\subsection{Evaluation Results} 
To evaluate the effectiveness of our approach, we calculate the coverage and accuracy. Because, a rule R can be assessed by it's coverage and accuracy \cite{han2011data}. Given a tuple, x, from a class labeled dataset, D, let $n_{covers}$ be the number of tuples covered by R;  $n_{correct}$ be the number of tuples correctly classified by R; and $|D|$ be the number of tuples in D. According to \cite{han2011data}, we can define the coverage and accuracy of R as -

\begin{equation}
	Coverage = \frac{n_{covers}}{|D|}
\end{equation}	
\begin{equation}
	Accuracy = \frac{n_{correct}}{n_{covers}}
\end{equation}

\begin{table}[htbp]
	\centering
	\caption{Accuracy and Coverage for different confidence threshold}
	\label{user-XY}
	\begin{tabular}{|c|c|c|c|c|} 
		\hline
		\bf User & \bf Confidence & \bf Accuracy & \bf Coverage \\  
		\hline
		& 100\% & 1.00 & 0.54 \\ 
		\cline{2-4}
		User U1 & 80\% & 0.93 & 0.59 \\ 
		\cline{2-4}
		& 60\% & 0.82 & 0.65 \\ 
		\hline
		& 100\% & 1.00 & 0.62 \\ 
		\cline{2-4}
		User U2 & 80\% & 0.90 & 0.67 \\ 
		\cline{2-4}
		& 60\% & 0.80 & 0.78 \\ 
		\hline
	\end{tabular}
\end{table}

In this experiment, we show the effect of confidence on coverage and accuracy of the discovered rules produced by our approach. Table \ref{user-XY} shows the results for the user U1 and U2 respectively for different confidence thresholds. In addition to this, Figure \ref{fig:Dataset-2} shows the average results of coverage and accuracy for different confidence thresholds from 100\% (maximum) below to 60\% (lowest) utilizing all the datasets. As confidence is associated to \textit{rules' accuracy}, we are not interested to take into account below 60\% as confidence preference.

If we observe the results in Table \ref{user-XY} and Figure \ref{fig:Dataset-2}, we can see that for 100\% confidence threshold, the accuracy is high but the coverage is low. The reason is that using 100\% confidence threshold, our approach outputs those rules, in which the user is always unavailable to answer a phone call. As this maximum threshold is satisfied by less number of rules, the approach then produces fewer rules with this high confidence value and as a result coverage decreases. On the other hand, if the confidence threshold becomes low, the accuracy decreases as it associates with the confidence but the coverage increases. Because it is then satisfied by more number of rules and outputs all the rules above this threshold. As individuals' preferences may differ in the real world, we allow users to configure the accuracy-coverage trade off based on their individual preferences (e.g., say 80\%). 

\section{Conclusion and Future Work}
\label{Conclusion and Future Work}
In this paper, we have presented an approach to infer the opportune moments for phone call interruptions based on user's unavailability, i.e., when a user is unable to answer the incoming phone calls. We also extract the corresponding phone silent mode configuring rules by analyzing the phone call activity patterns of individual mobile phone users. Experiments on real mobile phone datasets show that our approach is effective to identify the opportune moments for call interruptions and generates corresponding silent mode configuring rules by capturing the dominant activity of individuals at various times-of-the-day and days-of-the-week, according to individuals' preferences. We believe that our approach opens a promising path for future research on data-driven intelligent context-aware systems utilizing phone log data.

In future work, we plan to develop an intelligent phone call interruption management system that automatically configures silent notification mode based on the rules extracted by our approach, in order to provide the personalized services by minimizing the interruptions for the end mobile phone users.

\bibliographystyle{plain}
\bibliography{bibfile/PhoneSilent}
\end{document}